\documentclass[pre,letterpaper,floatfix,superscriptaddress]{revtex4}
\usepackage{graphicx,amsmath,amssymb}
\begin{document}
\title{Spatial fluctuations of a surviving particle in the trapping reaction}
\author{L. Anton}
\email{lucian.anton@infim.ro} \affiliation{Department of Physics
and Astronomy, University of Manchester, M13 9PL, U.K.}
\affiliation{Institute of Atomic Physics, INFLPR, Lab 22, PO Box MG-36
R76900, Bucharest, Romania}
\author{R. A. Blythe}\email{richardb@theory.ph.man.ac.uk}
\affiliation{Department of Physics and Astronomy, University of
Manchester, M13 9PL, U.K.}
\author{A. J. Bray} \email{bray@theory.ph.man.ac.uk}
\affiliation{Department of Physics and Astronomy, University of
Manchester, M13 9PL, U.K.}

\date{\today} 

\begin{abstract}
We consider  the trapping reaction, $A+B\rightarrow B$,  where $A$ and
$B$  particles have  a diffusive  dynamics characterized  by diffusion
constants $D_A$ and  $D_B$. The interaction with $B$  particles can be
formally incorporated in an effective dynamics for one $A$ particle as
was recently  shown by Bray  {\it et al}.\  [Phys.\ Rev.  E  {\bf 67},
060102  (2003)]. We  use this  method to  compute, in  space dimension
$d=1$, the  asymptotic behaviour of the  spatial fluctuation, $\langle
z^2(t)\rangle^{1/2}$, for a surviving $A$ particle in the perturbative
regime,  $D_A/D_B\ll  1$,  for   the  case  of  an  initially  uniform
distribution of $B$  particles.  We show that, for  $t\gg 1$, $\langle
z^2(t)\rangle^{1/2} \propto  t^{\phi}$ with $\phi=1/4$.   By contrast,
the  fluctuations of  paths constrained  to return  to  their starting
point  at time  $t$ grow  with the  larger exponent  $1/3$.  Numerical
tests are consistent with these predictions.
\end{abstract}

\maketitle

\section{Introduction}
Reaction-diffusion   problems   are   a   paradigm   of   irreversible
nonequilibrium   dynamics  \cite{Redner}.    Among   the  conceptually
simplest of such problems is the  trapping reaction, $A + B \to B$, in
which $A$  and $B$ particles  diffuse in space.   When an $A$  and $B$
particle  meet the  $A$ particle  is annihilated,  while there  are no
interactions between  particles of the same species.  Though simple to
state,  this problem  is  quite subtle:  for  example, the  asymptotic
behaviour  of the  $A$-particle density  for one-  and two-dimensional
systems has only recently been computed \cite{BB,BB03}.

Recently       there       has       been       renewed       interest
\cite{BB02,OBMC,MOBC1,MOBC2,MOBC3,BMB,MB,YA}    in    this   reaction,
stimulated  in  part by  the  advent  of  both new  numerical  methods
\cite{MG}  and  new analytical  techniques  \cite{BB,BB03}.  The  main
focus  of   interest  in  this  system  has   traditionally  been  the
computation  of  the   time-dependence  of  the  $A$-particle  density
\cite{BL}. Note  that, since  the $A$-particles are  independent, this
computation  is   equivalent  to  the  computation   of  the  survival
probability, $Q(t)$, of  a single $A$-particle moving in  a sea of $B$
particles. In the case of one  space dimension, to which this paper is
devoted,  $Q(t)$ is  known to  decay asymptotically  as $\exp(-\lambda
t^{1/2})$   \cite{BL}.   Recently,  by   a  combination   of  bounding
arguments,   the    constant   $\lambda$   was    determined   exactly
\cite{BB,BMB}: $\lambda  = (2/\pi)\rho(4\pi D_B)^{1/2}$,  where $\rho$
and $D_B$ are  the density and diffusion constant  respectively of the
$B$-particles.  Note that the constant $\lambda$ does {\em not} depend
on  the  $A$-particle  diffusion  constant,  $D_A$,  and  the  leading
asymptotic result is the same as for a stationary $A$-particle.

A parallel  advance in algorithmic methods \cite{MG}  has enabled very
long   simulations  of   the  model,   especially  in   one  dimension
\cite{MG,BB03},  in which  $Q(t)$ can  be made  arbitrarily  small (of
order $10^{-70}$, say, in  reasonable computer time).  Remarkably, the
predicted asymptotic  decay is not observed in simulations even for
these  very  small  values  of  $Q(t)$,  motivating  a  study  of  the
preasymptotic behaviour.  In a  companion paper \cite{Anton04}, we have
employed  a path-integral  approach \cite{BMB}  to compute,  for small
ratio $D_A/D_B$ of diffusion  constants, the leading correction to the
asymptotic form  quoted above.   Including this term  gives reasonable
agreement with data.

In the present paper we address, within the path-integral formalism, a
different aspect  of the problem.  Specifically, we  study the spatial
fluctuations, $\langle  x^2(t) \rangle$, of  the $A$-particle averaged
over  the  surviving  trajectories  at  time $t$.  We  find  that  the
fluctuations  are   subdiffusive:  $\langle  x^2(t)\rangle^{1/2}  \sim
t^{\phi}$  with  $\phi<1/2$.   In  their  numerical  work,  Mehra  and
Grassberger \cite{MG}  find $\phi$  to be in  the range $0.25  - 0.3$,
where the effective exponent seems to be decreasing at larger $t$.  We
will argue that $\phi=1/4$. 

Our approach is  based on a recent development  in ref.~\cite{BMB}. The
main idea is  to consider the probability of  intersection before time
$t$, $P(x,t;z)$, between a $B$ particle starting from $x$ at $t=0$ and
a fixed trajectory, $z(t)$, of the $A$ particle starting at $x=0$ when
$t=0$. This probability  can be averaged over the  initial position of
the $B$ particle  (assumed uniform in length $V$,  where we specialize
to  space dimension $d=1$):  $R(t;z)/V =  (1/V)\int dx  P(x,t;z)$. The
survival probability, i.e.\ the probability of no collision up to time
$t$, for  the trajectory $z$ of  the $A$ particle  is $1-R(t;z)/V$. If
the system has $N$  $B$-particles, uniformly distributed in the length
$V$,   the  survival   probability   for  the   given  trajectory   is
$(1-R(t;z)/V)^N$  which,  in the  continuum  limit  $N \to\infty$,  $V
\to\infty$, with $\rho=N/V={\rm constant}$, becomes $\exp\{-\mu[z]\}$,
where $\mu[z]=\rho R(t;z)$.

The  statistical properties  of the  $A$ particle are calculated by
averaging over  the trajectories, where each  trajectory is weighted
with its survival  probability multiplied by its Brownian weight. In 
this way one obtains the action functional \cite{BMB}
\begin{equation}
S[{z}] = \frac{1}{4D_A}\int_0^t d\tau\,
\left(\frac{d{z}}{d\tau}\right)^2 + \mu[{z}],
\end{equation}  
which  defines  an  effective  dynamics  for  surviving  $A$  particle
trajectories \cite{BMB}.

Further progress requires an explicit expression for the functional
$\mu[z]$. In ref.~\cite{BMB} it was  shown that the
constant trajectory, $z(t)=0$, gives the global minimum of $\mu[z]$,
and furthermore  a systematic expansion in $D_A/D_B$  was obtained for
$\mu[z]$. To first order one obtains \cite{BMB}
\begin{equation} \label{eq:mu1}
  \mu[z] = \lambda t^{1/2}+
  \frac{\lambda}{8\pi D_B}\int_{0}^{t}
  \frac{dt_1}{(t-t_1)^{1/2}}\int_{0}^{t_1}
  \frac{dt_2}{t_{2}^{1/2}(t_1-t_2)^{3/2}}
  [z(t_1)-z(t_2)]^2 + O(z^4)\ ,
\end{equation}
where $\lambda=(2/\pi)\rho(4\pi D_B)^{1/2}$.

In  this paper  we  study  the asymptotic  properties  of the  spatial
fluctuation of  the $A$ particle, using the  formalism described above
in    the   quadratic    approximation   for    $\mu[z]$    given   by
Eq. \eqref{eq:mu1}.  The  paper is laid out as  follows. In section II
we  apply  the  formalism  developed  in  refs.~\cite{BMB,Anton04}  to
compute  the root-mean-square  fluctuation of  surviving trajectories,
and show  that it scales  as $t^{1/4}$. In  section III we  adapt this
formalism to  compute the  root-mean-square fluctuation at  time $t/2$
for  surviving `loop'  trajectories,  constrained to  return to  their
starting points at  time $t$. This second calculation  is motivated by
the fact that it involves a very different mathematical treatment, and
it is not obvious that the same fluctuation exponent will be obtained.
Indeed,  we  find that  this  second  quantity  scales with  a  larger
exponent  $1/3$,  which  seems  at  first sight  counter  to  physical
intuition.  Numerical  results presented in section  III are, however,
consistent  with these  two different  predictions.  Section  IV  is a
summary  and conclusion. Some  technical details  are relegated  to an
appendix.

\section{Spatial fluctuations of surviving trajectories}

The probability  for the  $A$ particle, starting  from the  origin and
following a trajectory $z(t)$, to reach $x$ at time $t$ can be written
as
\begin{equation}\label{eq:pxt}
p(x,t)=\mathcal{N}\int_{z(0)=0}^{z(t)=x}\mathcal{D}[z(t)]\exp(-S[{z}])\ .
\end{equation}
Since we are  interested in the spatial fluctuations  of the surviving
trajectories we choose the normalization factor $\cal{N}$ such that
\begin{align}
  \int dx\, p(x,t)=1\ .
\end{align}
We can extract the $x$-dependence of the functional integral, Eq.\   
\eqref{eq:pxt}, by expanding the trajectories $z(t)$ around the 
`classical path', $z_{cl}(t)$, which minimizes the action $S$. 
Since the action is a quadratic functional of $z(t)$, we have
\begin{equation}
   p(x,t)=\mathcal{N}\exp(-S[{z_{cl};x,t}])
   \int_{z(0)=0}^{z(t)=0}\mathcal{D}[z(t)]\exp(-S[{z}])\ ,
\end{equation}
and the remaining functional integral, over the fluctuations around 
the classical path, can be absorbed into the normalization constant, 
as can the $z$-independent part of $\mu[z]$, see Eq. \eqref{eq:mu1}.
  
To  compute the  probability distribution  $p(x,t)$, therefore,  it is
sufficient  to find  the classical  path $z_{cl}$,  i.e.\ the  path of
least  action. The  analysis is  simplified by  the introduction  of a
dimensionless   action,   $\tilde{S}$,   through   the   substitutions
$t_1\rightarrow  tu$,  $t_2\rightarrow  tv$,  $z(t)\to  x\xi$  in  Eq.
\eqref{eq:mu1}. This gives
\begin{equation}\label{eq:dimlessaction}
\tilde{S} \equiv  \frac{2D_At}{z^2}S=\frac{1}{2} \int_{0}^{1}
  du\,{\dot\xi(u)}^2+\frac{1}{2} \gamma(t) 
  \int_{0}^{1}\frac{du}{\sqrt{1-u}}\int_{0}^{u}
  \frac{dv}{v^{1/2}(u-v)^{3/2}}(\xi(u)-\xi(v))^2\ ,
\end{equation}
where   $\gamma=(\lambda/2\pi)(D_A/D_B)t^{1/2}=(\pi^2/2)g(t)$   is   a
dimensionless  constant,  the final  equality  defining  $g$. In  what
follows, it  is important  to bear in  mind that $g  \propto t^{1/2}$,
and that we are interested in the large-$t$ behaviour of the theory.

The  minimum action  can be  computed using  the observation  that any
trajectory  starting at  $\xi(0)=0$ and  ending at  $\xi(1)=1$  can be
written as  a straight line connecting  the end points  plus a Fourier
expansion for the deviation from this straight line,
\begin{equation}
  \xi(u) = u+\sum_{n\ge 1} a_n\sin(n\pi u)\ ,
\end{equation}
where the linear part $u$ can also be written as a Fourier sine series: 
\begin{equation}
  u = 1-\frac{2}{\pi}\sum_{n\ge 1} \frac{\sin(n\pi u)}{n}.
\end{equation}
With these substitutions we can express the action,
Eq. (\ref{eq:dimlessaction}), in terms of the matrix elements $A_{mn}$ 
introduced in \cite{Anton04},
\begin{align}
   A_{mn}&=\frac{1}{mn}\int_{0}^{1}\frac{du}{\sqrt{1-u}}\int_{0}^{u}
  \frac{dv}{\sqrt{v}(u-v)^{3/2}}\notag\\&\times [\sin(n\pi
  u)-\sin(n\pi v)]\,[\sin(m\pi u)-\sin(m\pi v)].
\label{Amn}
\end{align}
After a straightforward computation we obtain:
\begin{equation}
  \tilde{S} =\frac{1}{2} + g\sum_{m,n}A_{mn}-\pi
  g\sum_{m,n}A_{mn}na_n+\frac{\pi^2}{4}\sum_{m,n}(n^2\delta_{mn}
  +gA_{mn}nm)a_ma_n.
\end{equation}
The classical path is obtained by minimizing this expression with 
respect to the coefficients $a_n$, which gives
\begin{equation}\label{eq:statsol}
  \tilde{a}_n=\frac{2}{\pi}g\sum_{m,q}\Bigl (
  \frac{1}{I+gA}\Bigr)_{nm}A_{mq}\ ,
\end{equation}
where we have made the rescaling $a_n =  \tilde{a}_n/n$.  Hence the 
minimum action can be expressed in terms of the matrix elements 
$A_{mn}$:
\begin{align}
 \tilde{S}_{cl} =
 \frac{2D_At}{x^2}S_{cl}&=\frac{1}{2}+g\sum_{mn}A_{mn} - g^2
 \sum_{\substack{m,n\\p,q}}\Bigl(\frac{1}{I+gA}\Bigr)_{np}A_{pq}
 A_{mn} \nonumber \\
 &=\frac{1}{2}+g\sum_{m,n,q}\Bigl(\frac{1}{I+gA}\Bigr)_{mq}A_{qn}.
\label{eq:action1}
\end{align}
The desired probability distribution $p(x,t)$ has the Gaussian form 
\begin{equation}
p(x,t) \propto \exp(-S_{cl}) = \exp\left(-\frac{\tilde{S}_{cl}}{2D_A t}\,
x^2\right),
\label{open}
\end{equation}
with  variance   $\langle  x^2  \rangle   =  D_At/\tilde{S}_{cl}$.  We
emphasize at this point  that $\tilde{S}_{cl}$ depends on time through
$g$. Our next goal, therefore,  is to determine the $g$-dependence, at
large $g$,  of the minimum action given by Eq.\ (\ref{eq:action1}). To
this  end  we  employ  a  finite-size  scaling  method,  by  computing
numerically the  action $S_{cl}$ for  an $A$ matrix truncated  to size
$N$, and analyzing the dependence of $\tilde{S}_{cl}$ on $g$ and $N$.

To motivate the  particular finite-size scaling form we  use, we first
consider the extreme limits $g \to \infty$ and $g \to 0$ at fixed $N$.
For $g \to \infty$,  Eq.\ (\ref{eq:action1}) becomes $\tilde{S}_{cl} =
1/2  + \sum_{n=1}^N  1 =  N +  1/2$,  while for  $g \to  0$ we  obtain
$\tilde{S}_{cl}  = 1/2 +  g\sum_{n,m=1}^N A_{mn}  + O(g^2)$.  From the
definition (\ref{Amn}) of $A_{mn}$ it  is easy to show that the double
sum converges to  the value $\pi^3/6$ for $N \to  \infty$. Thus we can
write $\tilde{S}_{cl} - 1/2 = H(g,N)$, where $H(g,N) \to N$ for $g \to
\infty$ and $H(g,\infty) \to  (\pi^3/6)g$ for small $g$. This suggests
the finite-size  scaling form $\tilde{S}_{cl} - 1/2  = Nh(g/N)$, where
$h(x) \sim x$ for small $x$ so that the $N$-dependence drops out as $N
\to \infty$.

\begin{figure}
\begin{center}
  \includegraphics{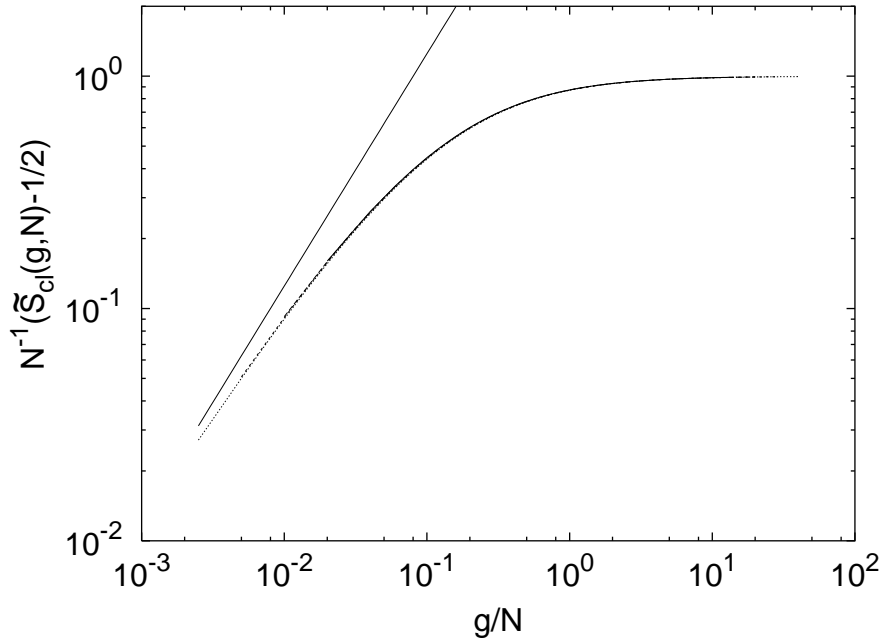}
    \caption{\label{fig:stat_action_all} Scaling of the $g$-dependent part
    of the dimensionless extremal action for free trajectories, for $N=50$, 
    $100$, $200$, $400$. 
    The continuous straight line with slope $1$ is a guide to the eye. It 
    indicates the expected asymptotic slope of the data for $g/N \ll 1$.  
    }
\end{center}
\end{figure}

A scaling  plot, testing  this form,  is presented in  Figure 1  for a
range of  values of $g$ and  $N$. The data collapse  is excellent (the
data sets are virtually  indistinguishable) confirming the validity of
the  assumed scaling  form.  We conclude  that,  in the  limit $N  \to
\infty$,  $\tilde{S}_{cl} \sim  g  \sim t^{1/2}$,  implying, via  Eq.\
(\ref{open}), that
\begin{equation}
\langle z^2(t)\rangle \propto  \rho^{-1}(D_{B}t)^{1/2}\ .
\end{equation} 
Note that  $\langle z^2(t)\rangle$ is independent of  $D_A$ at leading
order.

\section{Fluctuations of closed loops} 

We  turn now to  a discussion  of the  spatial fluctuations  of closed
loops, i.e.\ trajectories  that return to the origin  at time $t$. One
might  naively expect  that the  root-mean-square displacement  of the
trajectory  at,  say,  its   mid-point  $t/2$,  would  also  scale  as
$t^{1/4}$.  The mathematics  of this  calculation is,  however, rather
different and we shall see that the final result is different as well.

The probability distribution of the displacement $x$ of such a 
trajectory at time $t/2$ is given by
\begin{align}
  p^{(loop)}(x,t/2)=\mathcal{N}\int_{z(0)=0}^{z(t)=0} \mathcal{D}[z(t)]
  \delta(x-z(t/2)) e^{-S[z]}\ .
\end{align}
Applying the  standard `exponentiation' of  the delta function  via an
auxiliary integration,  representing $z$ by  a Fourier sine  series on
the interval $(0,t)$, and  performing the resulting Gaussian integrals
we obtain, after some straightforward algebra,
\begin{align}
  p^{(loop)}(x,t/2)\propto \exp\left[ -\frac{x^2}{2D_At}\,F(g) \right]
\label{loops}
\end{align}
 where
\begin{align}
  F(g) & = \frac{\pi^2}{4}
  \left[\sum_{m,n}\left(\frac{1}{1+gA}\right)_{mn}
  \frac{\sin(\pi m/2)\sin(\pi n/2)}{mn}\right]^{-1}\ .
\label{F}
\end{align}
It follows that 
\begin{equation}
\langle  x^2(t/2) \rangle_{loops} = \frac{D_At}{F(g)} = 
\frac{4D_At}{\pi^2}\sum_{m,n=1}^\infty\left(\frac{1}{1+gA}\right)_{mn}
  \frac{\sin(\pi m/2)\sin(\pi n/2)}{mn}\ . 
\end{equation}

Note the very different character  of the calculations of $p(x,t)$ and
$p^{(loop)}(x,t/2)$.   The  former  is  completely determined  by  the
classical path  connecting the  spacetime points $(0,0)$  and $(x,t)$.
The  fluctuations   around  this  path  are  not   important  for  the
calculation   of   $p(x,t)$  since   they   give  an   $x$-independent
contribution,  i.e.\  they  only   the  affect  the  normalization  of
$p(x,t)$,  which   can  in   any  case  be   fixed  by  hand   {\em  a
posteriori}. By contrast, the calculation of $p^{(loop)}(x,t/2)$, over
paths connecting the spacetime points $(0,0)$ and $(0,t)$, is entirely
concerned with fluctuations -- we need to explicitly integrate out the
fluctuations to determine $p^{(loop)}(x,t/2)$.

\begin{figure}
\begin{center}
  \includegraphics{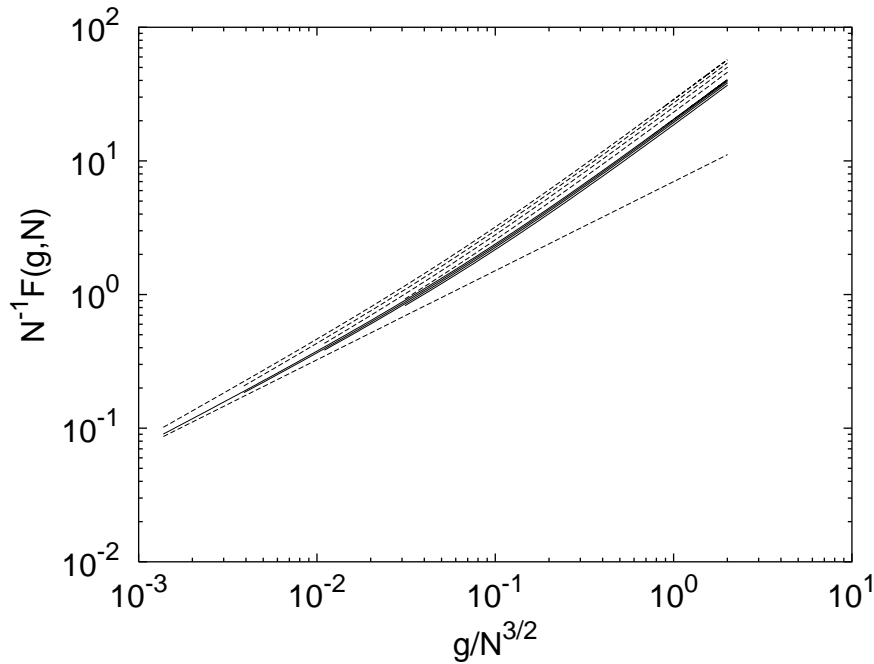}
    \caption{\label{fig:stat_action_loops} Scaling of the function 
     $F(g,N)$ (see text) for closed trajectories at time $t/2$, for 
    $N=50$, $100$, $200$, $400$: continuous curves -- full problem; broken 
    curves -- diagonal approximation. 
    The straight line with slope $2/3$ is a guide to the eye. It represents 
    the expected asymptotic slope of the data for $g/N^{3/2} \ll 1$. 
    }
\end{center}
\end{figure}

To determine $\langle x^2(t/2) \rangle_{loops}$ from Eq. (\ref{loops})
we need  to determine the form  of the function $F(g)$  for large $g$.
Again  we can exploit  a numerical  finite-size scaling  analysis.  In
view of  the considerations in  the preceding paragraph, we  expect a
rather different  finite-size scaling form for  the function $F(g,N)$,
obtained from truncating  the matrix A to size $N$,  than we found for
$\tilde{S}(g,N)$.   In  this  context  it  is helpful  to  recall  the
calculation \cite{Anton04} of the  subleading term in the $A$-particle
survival probability, which is also a fluctuation-dominated effect. In
that calculation it was found that approximating the $A$ matrix by its
diagonal  part gave  a  very  good guide  to  the finite-size  scaling
behaviour.  Using that  approach  here,  we are  led  to consider  the
function
\begin{equation}
F_{diag}(g,N) = \frac{\pi^2}{4}\left[\sum_{n=odd}^N
\left(\frac{1}{1+gA_{nn}}\right)\frac{1}{n^2}\right]^{-1}\ .
\label{Fdiag}
\end{equation}
For large $n$,  $A_{nn} \propto
n^{-3/2}$ (see  the appendix) and, for  large $N$ and $g$  it is clear
that large $n$  dominate the sum, justifying the use  of the large $n$
form of  $A_{nn}$. Inserting  this form in  Eq.\ (\ref{Fdiag})  we see
that, for $N \to \infty$,  $F_{diag} \sim g^{2/3} \propto t^{1/3}$ for
large   $g$,  whence  Eq.\   (\ref{loops})  gives   $\langle  x^2(t/2)
\rangle_{loops}        \sim        D_A        t/g^{2/3}        \propto
\rho^{-2/3}(D_AD_B)^{1/3}t^{2/3}$ within the diagonal approximation.

Furthermore,  the diagonal  approximation (\ref{Fdiag})  exhibits, for
$g$   and  $N$  both   large,  the   scaling  form   $F_{diag}(g,N)  =
g^{2/3}H(g/N^{3/2}) = Nh(g/N^{3/2})$. We  adopt this same scaling form
for   the  analysis   of  the   full  function   $F(g,N)$,  determined
numerically.  The  results, together with  the diagonal approximation,
are shown  in Figure 2. The  data collapse is  not as good as  for the
full set  of paths (Fig.~1), with  the data drifting  to higher values
for larger $N$.   The same is true, however, of  the diagonal data for
which  we know  by explicit  calculation that  the  assumed asymptotic
scaling form  is correct.  Indeed, the full  function $F(g,N)$ appears
to  be  converging  faster  with  increasing  $N$  than  the  diagonal
approximation  to it.  We  are therefore  confident  that the  assumed
scaling   is   asymptotically   correct   and,   as   a   consequence,
$\sqrt{\langle x^2(t/2)\rangle_{loops}} \sim t^{1/3}$.

An intriguing aspect  of this result is that  a subset of $A$-particle
trajectories have spatial fluctuations  scaling with a larger exponent
($\langle x^2 \rangle_{loops}^{1/2} \sim  t^{1/3}$) than the whole set
of  trajectories ($\langle  x^2 \rangle^{1/2}  \sim  t^{1/4}$).  While
this  result  seems at  first  sight  counterintuitive, the  numerical
scaling  analysis  strongly  suggests  that it  is  correct.   Further
supporting data will be presented in the following section.

It is  worth noting  that, in this  context, there is  nothing special
about the fluctuations of the mid-point of a closed path. For example,
we can  consider the  mean square fluctuation,  $\overline{\langle x^2
\rangle}_{loops}    =    t^{-1}\int_0^t    dt'\,    \langle    x^2(t')
\rangle_{loops}$, along  the path.  Using the  usual Fourier expansion
one readily obtains
\begin{equation}
\overline{\langle  x^2 \rangle}_{loops} = 
\frac{2D_At}{\pi^2}\sum_{n=1}^\infty \frac{1}{n^2} 
\left(\frac{1}{1+gA}\right)_{nn}\ .
\end{equation}
This quantity will also scale as $t^{2/3}$ (i.e.\ the root-mean square
fluctuation grows  as $t^{1/3}$) as  can be checked  explicitly within
the diagonal  approximation. As a  further twist, we can  consider the
fluctuations at  time $t/2$, or the root-mean  square fluctuation over
the whole path, of {\em all} trajectories that survive until time $t$.
These both  scale at $t^{1/3}$ in  contrast to the  fluctuation of the
end-point of the path which, as we have seen, grows as $t^{1/4}$.

That the fluctuations of the trajectory at time $t/2$, for paths which
survive to  time $t$,  could be larger  than the fluctuations  at time
$t/2$  of  paths which  survive  till  time  $t/2$ has  the  following
intuitive  explanation.  Trajectories  which survive  for a  long time
arise predominantly  from initial configurations of  the $B$ particles
in which the $A$ particle is initially in a large region devoid of $B$
particles.  If  the $A$  particle is to  survive until time  $t$, this
region is larger  than if it only has to survive  until time $t/2$, so
the  fluctuations in its  position at  time $t/2$  are expected  to be
larger in the former case than in the latter.

\section{Numerical Simulations}

Numerical  simulation data  were obtained  for both  the unconstrained
average   $\langle{z^2(t)}\rangle^{1/2}$  and  the   average  $\langle
z^2(t/2) \rangle_{loops}$ corresponding to surviving trajectories that
return to  the origin at time  $t$. The data were  generated using the
lattice-based algorithm of Mehra and Grassberger \cite{MG,BB03}, which
is  readily  adapted to  the  constrained  case.  The basic  algorithm
calculates  the survival probability,  averaged over  all $B$-particle
initial   conditions  and  trajectories,   of  a   given  $A$-particle
trajectory.   Final results  are obtained  by averaging  over  a large
number  (typically  $10^5$)   of  $A$-particle  trajectories  randomly
generated from the $2^t$ possible paths of length $t$. In the modified
algorithm, $A$-particle  trajectories are generated  dynamically. At a
given time,  the next  step is  taken to the  left with  a probability
equal  to the  number  of paths  that  reach the  origin  at time  $t$
(`return paths'), given that the first step is to the left, divided by
the  total number  of return  paths  from that  point. This  procedure
generates the ensemble of closed paths with the required weights.

The parameter values used  were $D_A=D_B=1/2$ and $\rho=1/2$.  In both
cases  we divide  the root-mean-square  value of  the  displacement by
$t^{1/4}$ and by  $t^{1/3}$ to test the different  predictions for the
two types of average.  The data are presented in Figures 3 and 4.  For
ease of  comparison, we have defined the end-time of the unconstrained 
paths to be $t/2$, i.e.\ all data are taken at the same real time.

\begin{figure}
\begin{center}
  \includegraphics{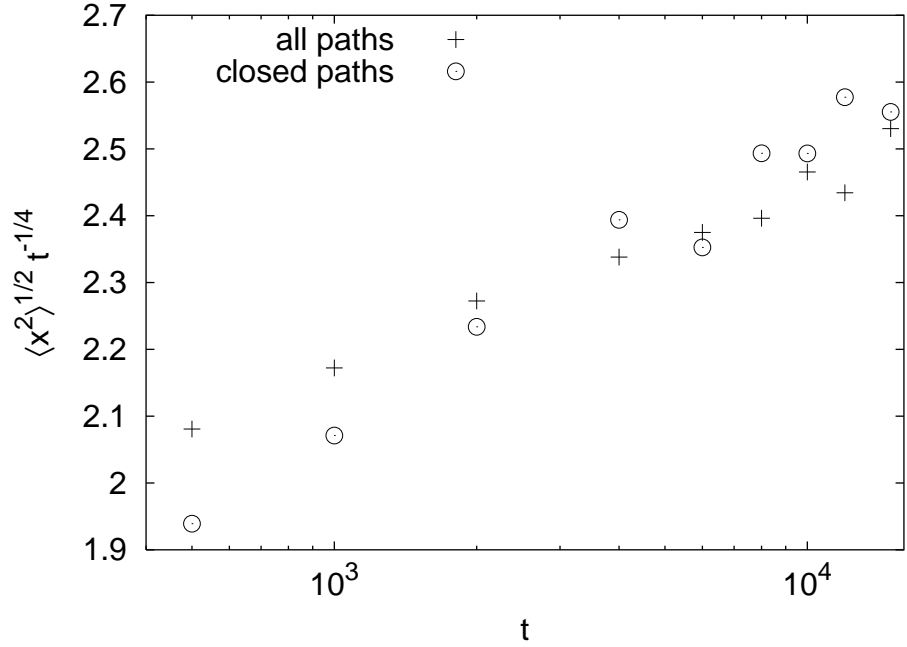}
    \caption{\label{fig:fluctuation1d4} Root mean square displacement 
    at time $t/2$, divided by $t^{1/4}$, for all paths and for closed  
    paths that survive till time $t$.  
    }
\end{center}

\end{figure}

\begin{figure}
\begin{center}
  \includegraphics{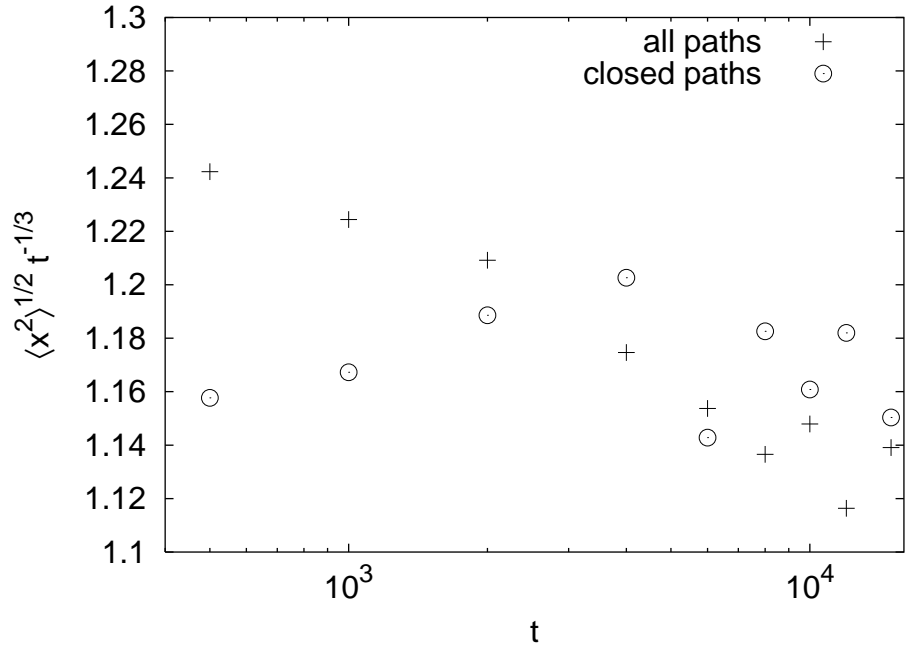}
    \caption{\label{fig:fluctuation1d3} Root mean square displacement 
    at time $t/2$, divided by $t^{1/3}$, for all paths and for closed  
    paths that survive till time $t$.  
    }
\end{center}

\end{figure}

Looking at Figure  3, we see that the closed-path  (loop) data show an
increasing trend with time $t$ consistent with an exponent larger that
$1/4$. A linear  fit on a log-log plot gives  an exponent $\phi \simeq
0.33$, consistent with the  predicted value of $1/3$. The unconstrained
data also show an increasing trend, but there is a suggestion that the
slope is decreasing at later times -- the  prediction  $\phi=1/4$  
requires that these data eventually reach a plateau as $t \to \infty$. 
A  linear fit, on  a log-log plot, gives $\phi \simeq 0.30$ over this 
range. The companion Figure 4,  in which the data are scaled by
$t^{1/3}$, tells  a similar story. The closed-path  data are consistent
with $\phi=1/3$, but the data including all paths show a clear decrease 
with time, indicative of a smaller exponent.

\begin{figure}
\begin{center}
  \includegraphics{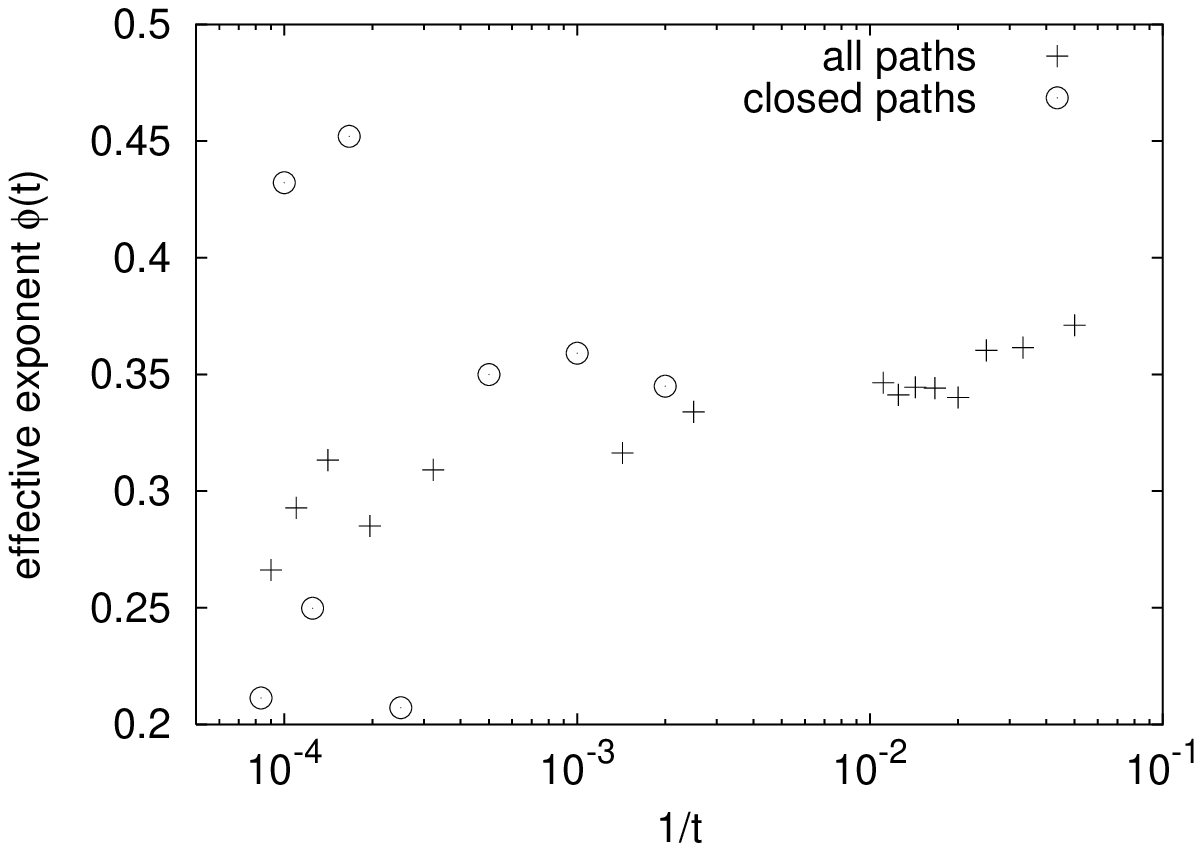}
    \caption{\label{fig:eff_exponent} Effective exponent $\phi(t)$ 
    for all paths and for closed paths, plotted against $1/t$. 
    }
\end{center}
\end{figure}

To  explore  this  behaviour  in  more  detail,  we  have  computed  a
time-dependent effective  exponent, $\phi(t)$. For  the computation we
use  the full data  set, of  which only  a small  subset (of  the `all
paths' data) is displayed in Figs.\  3 and 4.  The total time interval
is divided into consecutive bins  of size $dt_i$. The average value of
$\langle{z^2(t)}\rangle^{1/2}$ is  computed in each bin  to smooth the
data prior to taking  a numerical derivative.  The effective exponent,
$\phi(t_i)$, at  time $t_i$ is approximated by  the difference between
the logarithms of the average value  for bins $i$ and $i+1$ divided by
the difference of the logarithms  of the average positions of bins $i$
and  $i+1$. The  result  of this  procedure  is shown  in Fig.~5.  The
effective exponent is noisy, but is clearly decreasing with increasing
time  in a  manner consistent  with  the predicted  limiting value  of
$1/4$. The equivalent data for the closed paths is also shown. In this
case, however, there  are no more data available  beyond that shown in
Figures 3 and  4 -- the simulations are  computationally expensive due
to the  requirement that the paths  return to the origin  at time $t$.
As a result,  each bin contains a single data  point, and the effective
exponent is necessarily very noisy, but consistent with $\phi=1/3$.

\section{Conclusion}

We have studied the scaling  properties of the spatial fluctuations of
a  diffusing  particle interacting  with  diffusing  traps, using  the
effective action arising from eliminating the trap degrees of freedom.
The presence of  the trapping process renders the  fluctuations in the
position   of  surviving   particles  subdiffusive,   $\langle  z^2(t)
\rangle^{1/2} \propto t^{1/4}$, with  an amplitude that is independent
of  the $A$-particle diffusion  constant, $D_A$.   We have  also found
that the  fluctuations at  `half-time' of surviving  trajectories that
return  to the  origin at  time $t$  scale with  a  different exponent
$\langle  z^2(t/2)  \rangle_{(loops)}^{1/2}  \propto t^{1/3}$.   As  a
cautionary note we recall that  both of these results were obtained by
using the  quadratic approximation to the action  functional, which is
strictly valid  only in the  limit $D_A \ll  D_B$. It remains  an open
question to what extent corrections to the quadratic approximation can
modify the  results for general values  of $D_A/D_B$.  In  view of the
reasonable qualitative agreement of the theory with numerical data for
the  case  $D_A=D_B$,  however,  we  expect any  modifications  to  be
quantitative rather than qualitative.

\section*{Acknowledgements}

LA acknowledges support from the European Community Marie Curie 
Fellowship scheme under contract No. HPMF-CT-2002-01910. RAB 
acknowledges support from EPSRC under grant GR/R44768.

\appendix

\section{The matrix $A$}\label{sec:appendix}

The matrix element $A_{mn}$ is proportional to the double integral
\begin{equation}\label{eq:amninit}
  I_{mn}=mnA_{mn}=\int_{0}^{1}\frac{dx}{\sqrt{1-x}}\int_{0}^{x}
  \frac{dy}{\sqrt{y}(x-y)^{3/2}}[\sin(n\pi x)-\sin(n\pi y)]\,[\sin(m\pi
  x)-\sin(m\pi y)]
\end{equation}
which, by the variable substitutions $z=x-y$, $s=y$ and standard
manipulations of the trigonometric functions can be reduced to a sum of 
one-dimensional integrals, 
\begin{align}\label{eq:Imn1d}
  I_{mn}&=2\pi\cos\left(\frac{\pi(m-n)}{2}\right) \int_{0}^{1}
  \frac{dz}{z^{3/2}}\,\sin\left(\frac{n\pi z}{2}\right)
  \left(\sin \frac{m\pi z}{2}\right) 
  J_0\left(\frac{\pi}{2}(m-n)(1-z)\right )\notag \\ 
 &+2\pi\cos\left(\frac{\pi(m+n)}{2}\right) \int_{0}^{1}\frac{dz}{z^{3/2}}\,
  \sin\left(\frac{n\pi z}{2}\right)\sin\left(\frac{m\pi z}{2}\right) 
  J_0\left(\frac{\pi}{2}(m+n)(1-z)\right),
\end{align}
where $J_0(z)$ is the Bessel function of the first kind. The fact 
that $I_{mn}=0$ if $m+n=2p+1$ can be obtained directly from 
Eq. (\ref{eq:amninit}) on noticing that the kernel is symmetric on 
reflection about the line $y=1-x$.

For the case $m=n$, the dominant large-$n$ contribution comes from 
the first integral: 
\begin{equation}
  2\pi\int_{0}^{1} \frac{dz}{z^{3/2}}\,\sin^2\left(\frac{n\pi z}{2}\right) 
  \approx 2\pi
  \sqrt{n}\int_{0}^{\infty} \frac{dz}{z^{3/2}}\,
  \sin^2\left(\frac{\pi z}{2}\right)  = \sqrt{2}\pi^2 n^{1/2}\ .
\end{equation}

More elaborate is the calculation of the asymptotic behaviour for
$I_{mn}$ at large $n$ with $m$ fixed and large. We do the same
substitutions, $z=x-y$, $s=y$, in order to isolate the most singular
part in Eq. \eqref{eq:amninit}. Then we integrate once by parts
with respect to $z$ such that the singularity in $z$ becomes
integrable: $z^{-3/2}\rightarrow z^{-1/2}$. In the next step we
transform the products of trigonometric function into sums to give
\begin{align}
 \frac{1}{\pi}I_{nm}&=-(n-m)J_{n,n+m}-(n-m) J_{n-m,n-m} 
+(n+m) J_{n+m,n+m} +(n+m) J_{n,n-m} \nonumber \\ 
        & -(n+m)K_{n,n+m} - (n-m)K_{n-m,n-m} 
          + (n+m)K_{n+m,n+m} + (n-m) K_{n,n-m}\ ,
\end{align}
where the integrals $J_{n,m}$ and $K_{n,m}$ are given by
\begin{align}
  J_{n,m} & =\int_{0}^1\frac{dz}{z^{1/2}}\int_{0}^{1}
    \frac{ds}{s^{1/2}(1-s)^{1/2}}\sin\pi(nz+m(1-z)s) \\
  K_{n,m} & =\int_{0}^1\frac{dz}{z^{1/2}}\int_{0}^{1} \frac{ds
    (1-2s)}{s^{1/2}(1-s)^{1/2}}\sin\pi(nz+m(1-z)s)\ . 
\end{align}

Now we have to calculate these integrals to leading order in 
the $n \gg 1$, $m \gg 1$ and $n \gg m$. We show in detail
the computation for $J_{n,n+m}$. We deform the integral from 
$(0,1)$ to two parallel lines of the form $0 \rightarrow \pm i \infty$ 
and $ 1 \pm i\infty \to 1$, the sign being chosen such that the 
intermediate segment $\pm i \infty \to 1 \pm i\infty$ vanishes. 
This gives  
\begin{align}
  J_{n,n+m} &=\Im \int_{0}^1\frac{dz}{z^{1/2}}\int_{0}^{1}
    \frac{ds}{s^{1/2}(1-s)^{1/2}}e^{i\pi(nz+(n+m)(1-z)s)} \\ & = \Im
    \int_{0}^1\frac{dz}{z^{1/2}} e^{i\pi n z} \Biggl[ \int_{0}^{\infty}
    \frac{ i ds}{\sqrt{i} s^{1/2}(1-is)^{1/2}} e^{-\pi(n+m)(1-z)s} 
     \nonumber \\
    &- e^{i\pi\pi(n+m)(1-z)}\int_{0}^{\infty} \frac{ i ds}{\sqrt{-i}
    s^{1/2}(1+is)^{1/2}}e^{-\pi(n+m)(1-z)s}\Biggr] \\ & \approx \Im
    \Biggl[ \frac{\sqrt{i}}{\sqrt{(n+m)}}\int_{0}^{1} \frac{dz e^{i\pi
    n z}} {\sqrt{z(1-z)}}-\frac{e^{i\pi(n+m)}}{\sqrt{i}{\sqrt{(n+m)}}}
    \int_{0}^{1} \frac{dz e^{-i\pi m z}}{\sqrt{z(1-z)}} \Biggr] \\ &
    \approx \Im\Biggl[ \frac{\sqrt{i}}{\sqrt{(n+m)}} \Biggl(
    \frac{\sqrt{i}}{\sqrt{n}}-\frac{e^{i\pi n}}{\sqrt{i} \sqrt{n}}
    \Biggr) - \frac{e^{i\pi(n+m)}}{\sqrt{i}\sqrt{(n+m)}} \Biggl(
    \frac{\sqrt{-i}}{\sqrt{m}}+\frac{\sqrt{i} e^{-i\pi m}}{ \sqrt{m}}
    \Biggr) \Biggr] \\ & =\frac{1}{\sqrt{n+m}}\Bigl(\frac{1}{\sqrt{n}}
    - \frac{\cos\pi(n+m)}{\sqrt{m}}\Bigr)
\end{align}
where we have used, to leading order, 
\begin{equation}
  \int_{0}^{\infty} \frac{dx}{x^{1/2}}Q(x) e^{-kz}\approx Q(0)
  \frac{\sqrt{\pi}}{\sqrt{k}}
\end{equation}
when $Q(0)$ exists and $k\gg 1$.

Special attention must be paid to the point $z=1$ in the integral over
$s$,  as   the  argument  in   the  exponential  is  not   large  when
$z\rightarrow 1$ with $m$, $n$  fixed. However, one can show that this
region gives a subleading contribution.

\begin{figure}
  \includegraphics{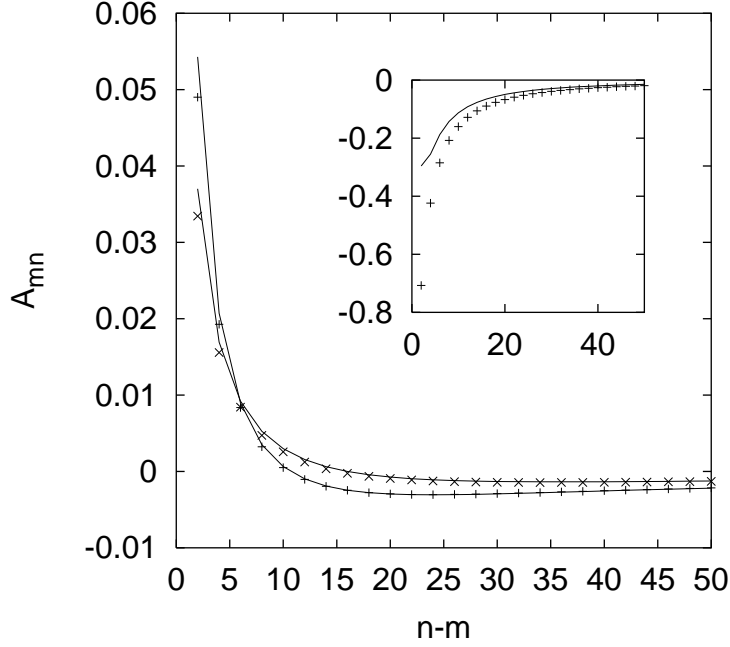}
    \caption{\label{fig:amnfit} $A_{mn}$ as s function of $n$ at $m=50,55$
    starting from the main diagonal of the matrix. In the inset $A_{1n}$
    is compared with the asymptotic result. The symbols are the from
    the numerical integration of Eq. \eqref{eq:amninit}, the lines are
    the asymptotic approximation, Eq. \eqref{eq:amnasympt}.}
\end{figure}

Applying the same  procedure to the remaining integrals,  we find that
the contribution of $J_{n-m,n-m}$, $J_{n+m,n+m}$ and $K_{n-m,n-m}$, 
$K_{n+m,n+m}$ cancel each other to leading order and we finally obtain 
the closed-form expression
\begin{equation}\label{eq:amnasympt}
  A_{mn}\approx
  \frac{2\pi}{mn}\Big( \frac{\sqrt{n}}{\sqrt{n-m}} -
  \frac{\sqrt{m}+\sqrt{n}}{\sqrt{m+n}} \Bigr)
\end{equation}
for $m\gg1$, $n\gg1$  and $n\gg m$. It is worth  noting that the above
result is a  good approximation for any value of  $m$, $n$ outside the
main diagonal as one can  see from Fig.~\ref{fig:amnfit}. This fact is
useful  in  numerical computations  since  the  integrals that  define
$A_{mn}$ have strongly oscillating integrands for large $m$, $n$.
 
\begin{thebibliography}{99}

\bibitem{Redner}
See, e.g., S. Redner in {\em Nonequilibrium Statistical Mechanics in One 
Dimension}, edited by V. Privman (Cambridge University Press, Cambridge, 
1997), and references therein. 

\bibitem{BB}
A. J. Bray and R. A. Blythe, 
Phys.\ Rev.\ Lett.\ {\bf 89} 150601, (2002).

\bibitem{BB03}
R. A. Blythe and A. J. Bray,
Phys.\ Rev.\ E {\bf 67}, 041101 (2003).

\bibitem{BB02}
R. A. Blythe and A. J. Bray, J. Phys.\ A {\bf 35}, 10503 (2002). 

\bibitem{OBMC}
G. Oshanin, O. Benichou, M.Coppey and M. Moreau, 
Phys.\ Rev.\ E {\bf 66}, 060101 (2002). 

\bibitem{MOBC1}
M. Moreau, G. Oshanin, O. Benichou and M. Coppey, 
Phys.\ Rev.\ E {\bf 67}, 045104 (2003). 

\bibitem{MOBC2}
M. Moreau, G. Oshanin, O. Benichou and M. Coppey,
Physica A {\bf 327}, 99 (2003).

\bibitem{MOBC3}
M. Moreau, G. Oshanin, O. Benichou and M. Coppey,
Phys.\ Rev.\ E {\bf 69}, 046101 (2004).  

\bibitem{BMB}
A. J. Bray, S. N. Majumdar and R. A. Blythe,
Phys.\ Rev.\ E {\bf 67}, 060102 (2003). 

\bibitem{MB}
S. N. Majumdar and A.J. Bray, 
Phys.\ Rev.\ E {\bf 68}, 045101 (2003).

\bibitem{YA}
S. B. Yuste and L. Acedo, 
Physica A {\bf 336}, 334-346 (2004).

\bibitem{MG} 
V. Mehra V and P. Grassberger, 
Phys.\ Rev.\ E {\bf 65} 050101 (2002). 

\bibitem{BL}
M. Bramson and J. L. Lebowitz, 
Phys.\ Rev.\ Lett.\ {\bf 61}, 2397 (1988). 

\bibitem{Anton04}
L. Anton and A. J. Bray, J. Phys. A {\bf 37}, 8407(2004).   

\end{thebibliography}
\end{document}